\documentclass{article} 
\usepackage{iclr2025_conference,times}

\usepackage{amsmath,amsfonts,bm}

\def\eqref#1{equation~\ref{#1}}

\def\1{\bm{1}}

\DeclareMathAlphabet{\mathsfit}{\encodingdefault}{\sfdefault}{m}{sl}
\SetMathAlphabet{\mathsfit}{bold}{\encodingdefault}{\sfdefault}{bx}{n}

\usepackage{hyperref}
\usepackage{url}
\usepackage{graphicx}

\title{Adiabatic Fine-Tuning of Neural Quantum States Enables Detection of Phase Transitions in Weight Space\thanks{Code available at: \url{https://gitlab.com/QMAI/papers/nqsweights}}}

\author{Vinicius Hernandes, Thomas Spriggs, Saqar Khaleefah \& Eliska Greplova \\
QuTech and Kavli Institute of Nanoscience\\
Delft University of Technology\\
Delft, Netherlands\\
\texttt{v.hernandes@tudelft.nl}}
\iclrfinalcopy 
\begin{document}
\maketitle

\begin{abstract}
Neural quantum states (NQS) have emerged as a powerful tool for approximating quantum wavefunctions using deep learning. While these models achieve remarkable accuracy, understanding how they encode physical information remains an open challenge. In this work, we introduce adiabatic fine-tuning, a scheme that trains NQS across a phase diagram, leading to strongly correlated weight representations across different models. This correlation in weight space enables the detection of phase transitions in quantum systems by analyzing the trained network weights alone. We validate our approach on the transverse field Ising model and the J1-J2 Heisenberg model, demonstrating that phase transitions manifest as distinct structures in weight space. Our results establish a connection between physical phase transitions and the geometry of neural network parameters, opening new directions for the interpretability of machine learning models in physics.
\end{abstract}

\section{Introduction and Related Work}

The simulation of quantum many-body systems represents one of the most challenging problems in computational physics, primarily due to the exponential growth of the Hilbert space with system size \citep{feynman2018simulating, nielsen2010quantum}. Neural Quantum States (NQS) have emerged as a promising approach, leveraging the expressive power of deep learning to represent quantum wavefunctions \citep{carleo2017solving, medvidovic2024neural}. In this approach, neural networks serve as ansätze for the wavefunction, mapping quantum states to their corresponding probability amplitudes. While these models give very good results in ground state energy calculations \citep{chen2023efficient, valenti2022correlation, wu2024variational, rende2024simple}, it remains challenging to understand how they encode physical information.

Recent work by Rende et al.~\citep{rende2024fine} introduced a fine-tuning strategy in which a network pretrained near a phase transition is adapted across a phase diagram by updating only the output layer, reducing computational costs while maintaining accuracy across phases. Building on this idea, we investigate how phase transitions relate to the geometry of neural network weight space. In deep learning, various studies have explored the structure of weight space with diverse objectives, ranging from analyzing the loss landscape to understand mode connectivity \citep{garipov2018loss, entezari2021role}, to leveraging neural network weights as data for downstream predictions \citep{eilertsen2020classifying, unterthiner2020predicting, schurholt2024towards}. Meanwhile, in quantum physics, phase transitions are traditionally characterized through order parameters and correlation functions. The intersection of these fields has led to emerging research on how neural networks capture physical symmetries and order parameters \citep{wetzel2020discovering, frohnert2024explainable, cybinski2024speak}, and how it's possible to exploit machine learning models to learn about physical phase transitions \citep{van2017learning, dawid2020phase, greplova2020unsupervised, arnold2021interpretable}. However, directly analyzing the weights of neural quantum states as a means of detecting phase transitions remains largely unexplored.

In this work, we introduce a fine-tuning scheme, called \textbf{adiabatic fine-tuning}, that enables systematic training of NQS across a phase diagram, leading to more accurate models, and highly correlated weights across different models. We show that this correlation can be leveraged to detect phase transitions by analyzing the PCA projection of trained network weights alone. Applying this method to the transverse field Ising model and the J1-J2 Heisenberg model, we demonstrate that phase transitions manifest as distinct geometric patterns in weight space. By directly analyzing the trained network weights, we show how the evolution of neural network parameters reflects physical features of quantum systems, providing a new method to understand phase transitions.

\section{Methodology}

\subsection{Neural Quantum State Training}

We trained neural networks models to approximate the quantum wavefunctions of the systems under study. Specifically, the wavefunctions for the transverse field Ising model (TFIM) and J1-J2 Heisenberg model were parameterized using neural networks, with the goal of minimizing the energy of the system. We implemented a restricted Boltzmann machine architecture with a single hidden layer, using system sizes of $N=8$ spins for the Ising model and $N=12$ spins for the J1-J2 model (see Appendix C for architectural details). In this study, we focus on small system sizes for which we can compute the exact wavefunction, allowing direct comparison to known results. 

\subsection{Transverse Field Ising Model}

The TFIM in one dimension is defined by the Hamiltonian:
\begin{equation}
\hat{H} = -J \sum_{i} \sigma_i^z \sigma_{i+1}^z - h \sum_{i} \sigma_i^x,
\end{equation}
where $J$ is the interaction strength between neighboring spins, and $h$ is the external transverse magnetic field. This model exhibits a quantum phase transition from a ferromagnetic to a paramagnetic phase at the critical field value $|h_c/J|= 1$

\subsection{J1-J2 Heisenberg Model}

The 1D J1-J2 model is described by the Hamiltonian:
\begin{equation}
\hat{H} = J_1 \sum_{i} \mathbf{S}_i \cdot \mathbf{S}_{i+1} + J_2 \sum_{i} \mathbf{S}_i \cdot \mathbf{S}_{i+2},
\end{equation}
where $\mathbf{S}_i$ represents the spin operator at site $i$, and $J_1$ and $J_2$ define the nearest and next-nearest neighbor interactions, respectively. This model exhibits several distinct magnetic phases depending on the ratio $J_2 / J_1$.

\subsection{Model Performance Metrics}

The performance of our model in obtaining the correct ground state is assessed by calculating the \textit{energy error} and \textit{infidelity}. The energy error, defined as \( |E_{\text{NQS}} - E_{\text{exact}}| \), quantifies the deviation of the neural network-estimated energy from the exact ground state energy. The infidelity, given by \( 1 - |\langle \psi_{\text{exact}} | \psi_{\text{NQS}} \rangle|^2 \), measures the discrepancy between the trained and exact wavefunctions, with lower values indicating better overlap.

\subsection{Fine-tuned Training Strategy}

To explore the structure of weight spaces across phase transitions, we employ two distinct training strategies. The first approach is \textbf{independent training}. In this method, each model is initialized randomly and trained separately for each value of the control parameter, $h$ (for the Ising model) or $J_2 / J_1$ (for the J1-J2 Heisenberg model). This strategy allows us to study how the model weights evolve when trained independently for each specific parameter setting. It serves as a baseline for understanding how weight space behaves when there is no continuity between parameter values.

The second approach we call \textbf{adiabatic fine-tuning}. In this case, we initialize each new model using the weights of a previously trained model from the neighboring point in the phase diagram. This ensures continuity in the evolution of model parameters, creating a smooth trajectory through weight space that reflects the gradual changes in the underlying quantum system. Fine-tuning allows us to capture the relationship between the geometry of weight space and physical phase transitions in a more connected and coherent manner. By comparing the results from fine-tuned training to those from independent training, we can identify phase transitions as distinct features in the neural network parameter evolution,  revealing the structural nature of weight spaces across these transitions.

This fine-tuned training strategy creates a connected trajectory in weight space, which resembles the concept of mode connectivity observed in deep learning. The continuity in parameter evolution provided by fine-tuning enables us to detect phase transitions in quantum systems as sharp, discernible features in the progression of the network parameters.

\subsection{Principal Component Analysis of Weights}

After training the models, we analyze the resulting weight vectors by performing principal component analysis (PCA). This dimensionality reduction technique identifies the dominant directions in weight space and helps reveal how the weights evolve as the quantum system's control parameters change.

\section{Results}

\subsection{Transverse Field Ising Model}

\begin{figure}[!h]
\begin{center}
\includegraphics[width=0.99\textwidth]{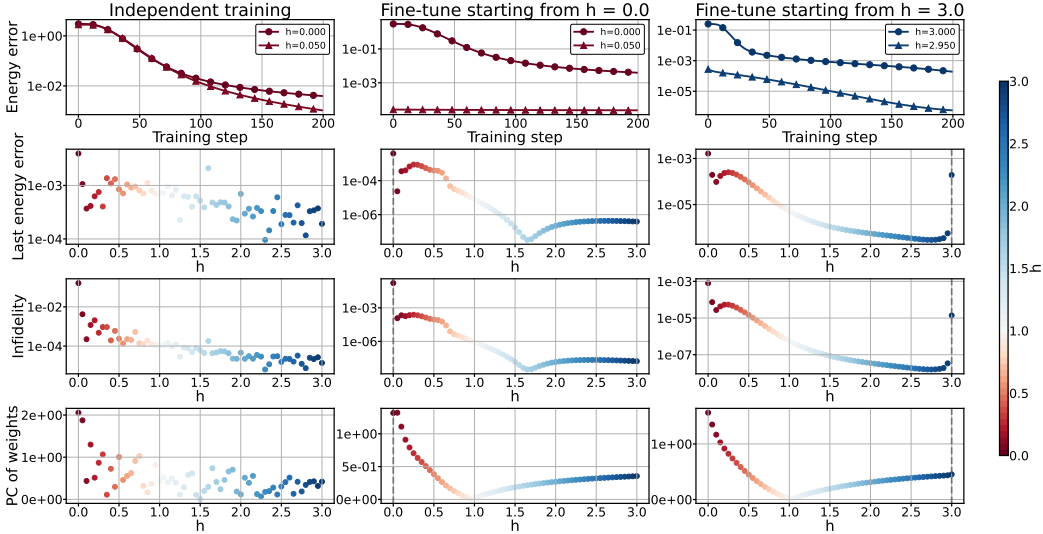}
\end{center}
\caption{Results for the Ising model. From top to bottom the rows show: the energy error over the first two training iterations, the last energy error values for each value of $h$, the infidelity values, and the principal component analysis of the weights. From left to right columns show the case of independent training, and fine-tuning models starting from $h=0.0$, and $h=3.0$. The colorbar represents the value of $h$.}
\label{fig:ising_results}
\end{figure}

Our analysis of the transverse field Ising model reveals a clear correspondence between the known quantum phase transition at $h_c = 1$ when $J=-1$ and features in the weight space trajectory. The first principal component shows a pronounced minimum at the critical point, providing a direct signature of the transition from the ferromagnetic to paramagnetic phase, as shown in Figure \ref{fig:ising_results}.

The energy plots in the top row of Figure \ref{fig:ising_results} reveal a key difference between independent training and fine-tuning strategies. In the case of independent training, each model starts at an arbitrarily high energy and gradually converges to a lower value. In contrast, fine-tuning ensures that the energy at the beginning of a new training run is already close to the final energy of the previous one. This continuity in the optimization process suggests that fine-tuned training preserves learned features from previous models, leading to a smoother transition through the phase diagram. Additionally, examining the second and third rows of the figure, we see that the final energy and infidelity values remain close for consecutive fine-tuned runs, except for lower values of the field $h$, where deviations become more pronounced.

\subsection{J1-J2 Heisenberg Model}

\begin{figure}[!h]
\begin{center}
\includegraphics[width=0.99\textwidth]{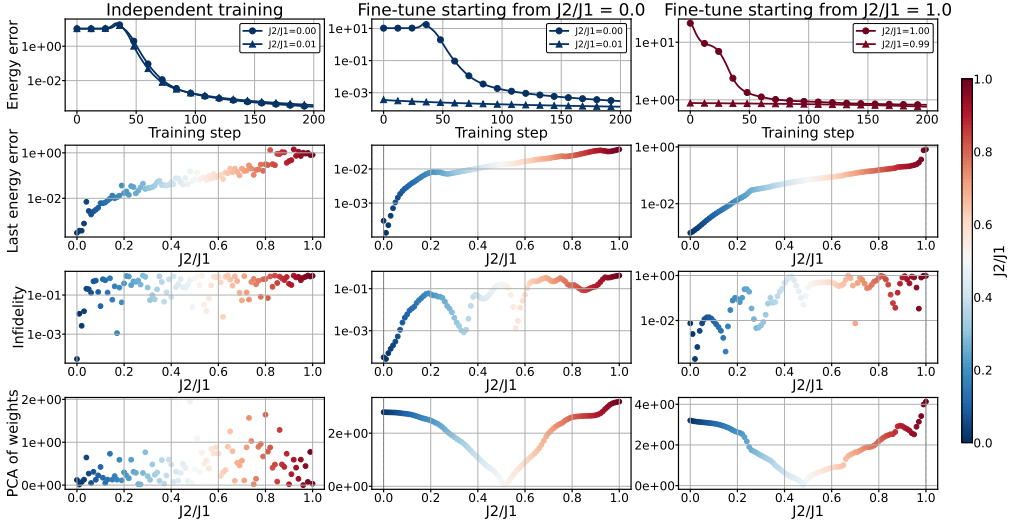}
\end{center}
\caption{Results for the J1-J2 model. From top to bottom the rows show: the energy error over the first two training iterations, the last energy error values for each value of $J2/J1$, the infidelity values, and the principal component analysis of the weights. From left to right columns show the case of independent training, and fine-tuning models starting from $J_2/J_1=0.0$, and $J_2/J_1=1.0$. The colorbar represents the $J_2/J_1$ ratio.}
\label{fig:j1j2_results}
\end{figure}

In the J1-J2 model, our method successfully identifies the change in the system's structure at $J_2/J_1 = 0.5$, appearing again as a minimum in the first PCA component, as shown in the bottom row of Fig.~\ref{fig:j1j2_results}.

The J1-J2 model exhibits the same trends in energy evolution, final energy values, and infidelity as observed in the Ising model, as shown in the top three rows of Figure \ref{fig:j1j2_results}. 

\section{Discussion and Conclusions}

Our work establishes a novel connection between phase transitions in quantum systems and the geometry of neural network weight spaces. The success of our method in detecting known phase transitions suggests that neural networks encode physically meaningful information in their weights in a structured and analyzable way.

This finding has several important implications. First, it provides a new tool for studying quantum phase transitions that doesn't require explicit construction of order parameters or prior knowledge of the relevant physical observables. Second, it offers insights into how neural networks encode physical information, this can be seen from the change in the behavior of the network's weights across the phase transition.

Future directions include extending this analysis to more complex quantum systems and investigating whether adiabatic fine-tuning  can be applied to other domains where neural networks model systems with phase transitions or structural changes. The connection to mode connectivity in deep learning also suggests potential applications in understanding the loss landscapes of neural networks more generally.


\bibliography{iclr2025_conference} \bibliographystyle{iclr2025_conference}
\appendix

\section{Energy for all models}

\begin{figure}[!h]
\begin{center}
\includegraphics[width=0.97\textwidth]{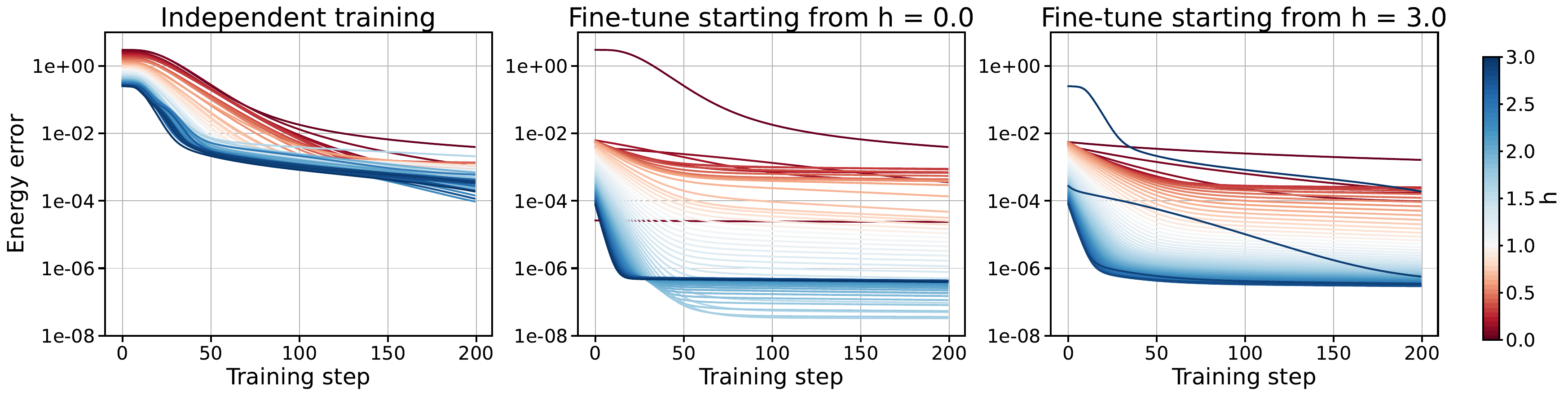}
\end{center}
\caption{Energy error as a function of training step for all values of $h$ in the TFIM model.}
\label{fig:ising_energies}
\end{figure}

\begin{figure}[!h]
\begin{center}
\includegraphics[width=0.99\textwidth]{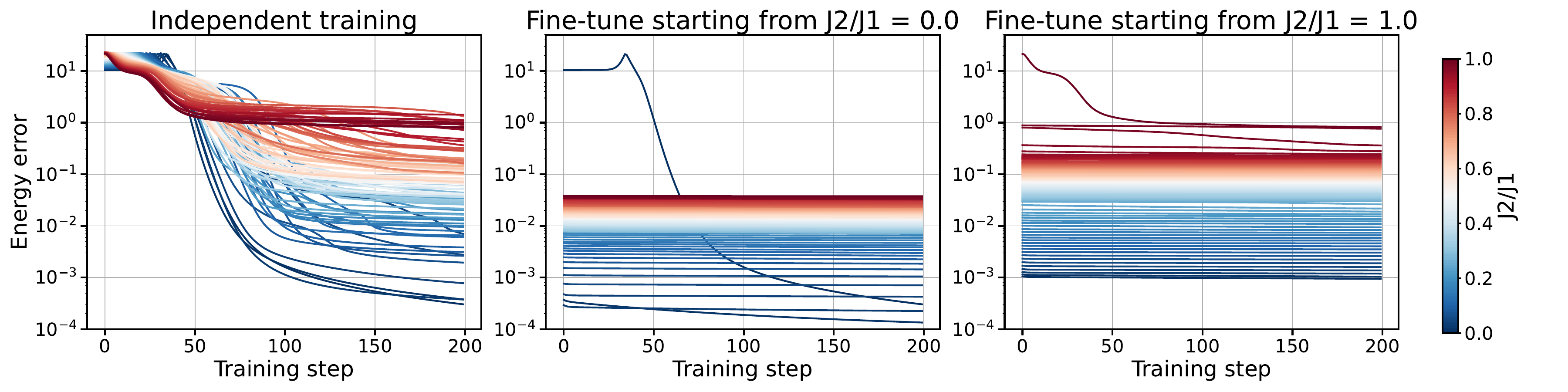}
\end{center}
\caption{Energy error as a function of training step for all values of $J2/J1$ in the J1-J2 model.}
\label{fig:j1j2_energies}
\end{figure}

Figures \ref{fig:ising_energies} and \ref{fig:j1j2_energies} show the energy error as a function of training steps for all values of the control parameters in both models. These plots highlight how the adiabatic fine-tuning strategy maintains lower initial energy errors compared to independent training, demonstrating the advantage of leveraging previously trained weights for improved convergence.

\section{PCA Visualizations}

\begin{figure}[!h]
\begin{center}
\includegraphics[width=0.99\textwidth]{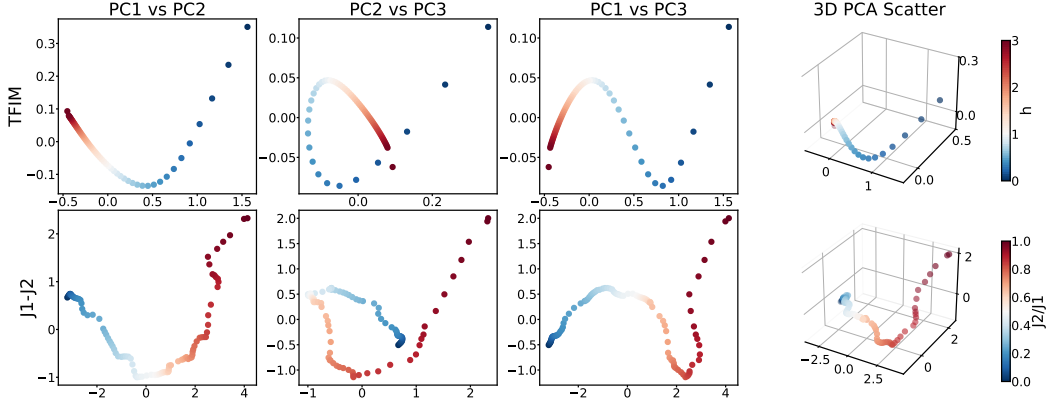}
\end{center}
\caption{Scatter plots and 3D visualizations of PCA projections of the weights of neural networks trained for different values of $h$ in the case of the Ising model (first row), and for different values of $J_2/J_1$ in the case of the J1-J2 model (second row). From left to right, columns show PC1 vs PC2 space, PC2 vs PC3 space, PC1 vs PC3 spaces, and the 3D PCA projection. The colorbars to the right indicate the corresponding parameter values for each model. The phase transition point for all images is represented by the divergence of the colormap, shown in white.}
\label{fig:pca_results}
\end{figure}

Further visualizations of the principal component analysis projections for both models are shown in Figure \ref{fig:pca_results}. The scatter plots and 3D visualizations provide a clearer understanding of how weights for models trained with different values of $h$ (Ising model) and $J_2/J_1$ (J1-J2 model) show strong correlation, forming a helix structure when projected in PC space. The existence of a similar pattern for different systems suggests that the fine tuning training scheme results in universal properties to be investigated in future work. 

\section{Network Architecture and Training Details}

\begin{table}[h]
\centering
\caption{Training hyperparameters}
\begin{tabular}{lcc}
\hline
Parameter & Ising Model & J1-J2 Model \\
\hline
Learning rate & 0.01 & 0.01 \\
Training steps & 200 & 200 \\
System size (spins) & 8 & 12 \\
Hidden layer ratio ($\alpha$) & 1 & 2 \\
Coupling increment ($\Delta h$ or $\Delta J_2/J_1$) & 0.025 & 0.01 \\
Coupling range & $[0,3]$ & $[0,1]$ \\
Network weights type & Real & Complex \\
\hline
\end{tabular}
\end{table}

Our neural network architecture consists of a restricted Boltzmann machine (RBM) with a single hidden layer using logcosh activation function, implemented as follows:

\begin{equation}
\psi(x) = \sum_i \text{logcosh}(W_ix + b_i)
\end{equation}

where $W_i$ represents the weights connecting the input to the hidden layer, and $b_i$ are the bias terms. The number of neurons in the hidden layer is determined by $\alpha N$, where $N$ is the number of input neurons (spins) and $\alpha$ is the hidden layer ratio. For the Ising model, we used real-valued weights, while for the J1-J2 model, we employed complex-valued weights (with PCA performed on the real components only). Hyperparameters used to trained models for both systems are shown in Table 1.

\end{document}